\documentclass[a4paper,11pt]{article}
\usepackage{pos}
\usepackage{multirow}
\usepackage[official]{eurosym}
\usepackage{setspace}
\usepackage{hyperref}

\title{Wavelength-shifting light traps for SWGO and other applications}

\author*[a]{M. Pihet}
\author[a,b]{M. Mariotti}
\author[a]{C. Arcaro}

\affiliation[a]{INFN Padua,\\
Via Marzolo, 8, 35131 Padua, Italy}

\affiliation[b]{University of Padua,\\
 Via VIII Febbraio, 2, 35122 Padua, Italy}

\onbehalf{for the SWGO collaboration\footnote{Complete list of SWGO authors and acknowledgments: \url{https://www.swgo.org/SWGOWiki/lib/exe/fetch.php?media=wiki:swgo\_collaboration\_icrc23.pdf}}} 

\emailAdd{marine.pihet@pd.infn.it}
\emailAdd{mose.mariotti@unipd.it}
\emailAdd{cornelia.arcaro@pd.infn.it}

\abstract{Wavelength-shifting (WLS) materials contain molecules that absorb light and reemit at longer wavelengths. They can be used for light detection because they provide a large effective area for low cost and they are able to efficiently trap and guide light because of total internal reflection processes. We are currently developing such a WLS detector, considering two main designs: A single-shift design with one wavelength shift (tile) and a double-shift design with two wavelength shifts (tile and fiber). As photodetectors we use small Silicon photomultipliers (SiPMs) with a high photon detection efficiency (PDE) and single-photon sensitivity. The double-shift layout goes at the expense of detection efficiency. In this design however, light is channeled to the two ends of a fiber, thus requiring a reduced photosensitive area compared to the single-shift layout. We will present the results of our measurements and show that light traps and SiPMs together represent a promising alternative to PMTs in case of a non-focused light beam. For the special case of SWGO, the application of light traps is also motivated by a possible improvement of the gamma/hadron separation, using a one-chamber tank with an array of wavelength-shifting light traps instead of a (two-chamber) tank with PMTs. Besides SWGO, new WLS detectors could also constitute useful and cheap technology for other experiments and use cases. The contribution summarizes our motivation and efforts to build a light trap detection module and to characterize its properties in terms of costs, temporal performance and detection efficiency.}

\ConferenceLogo{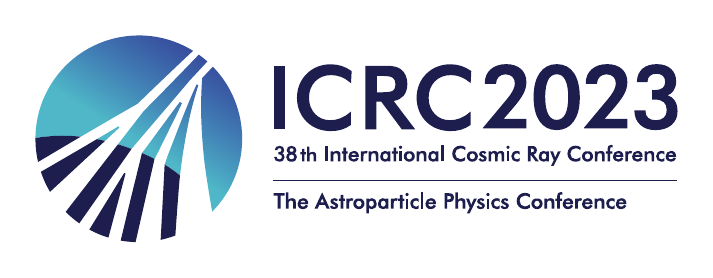}

\FullConference{%
38th International Cosmic Ray Conference (ICRC2023)\\
  26 July - 3 August, 2023\\
  Nagoya, Japan}

\begin{document}
\maketitle

\section{Introduction} \label{sec:introduction}

\begin{figure}[h]
    \centering
    \includegraphics[width=0.7\textwidth]{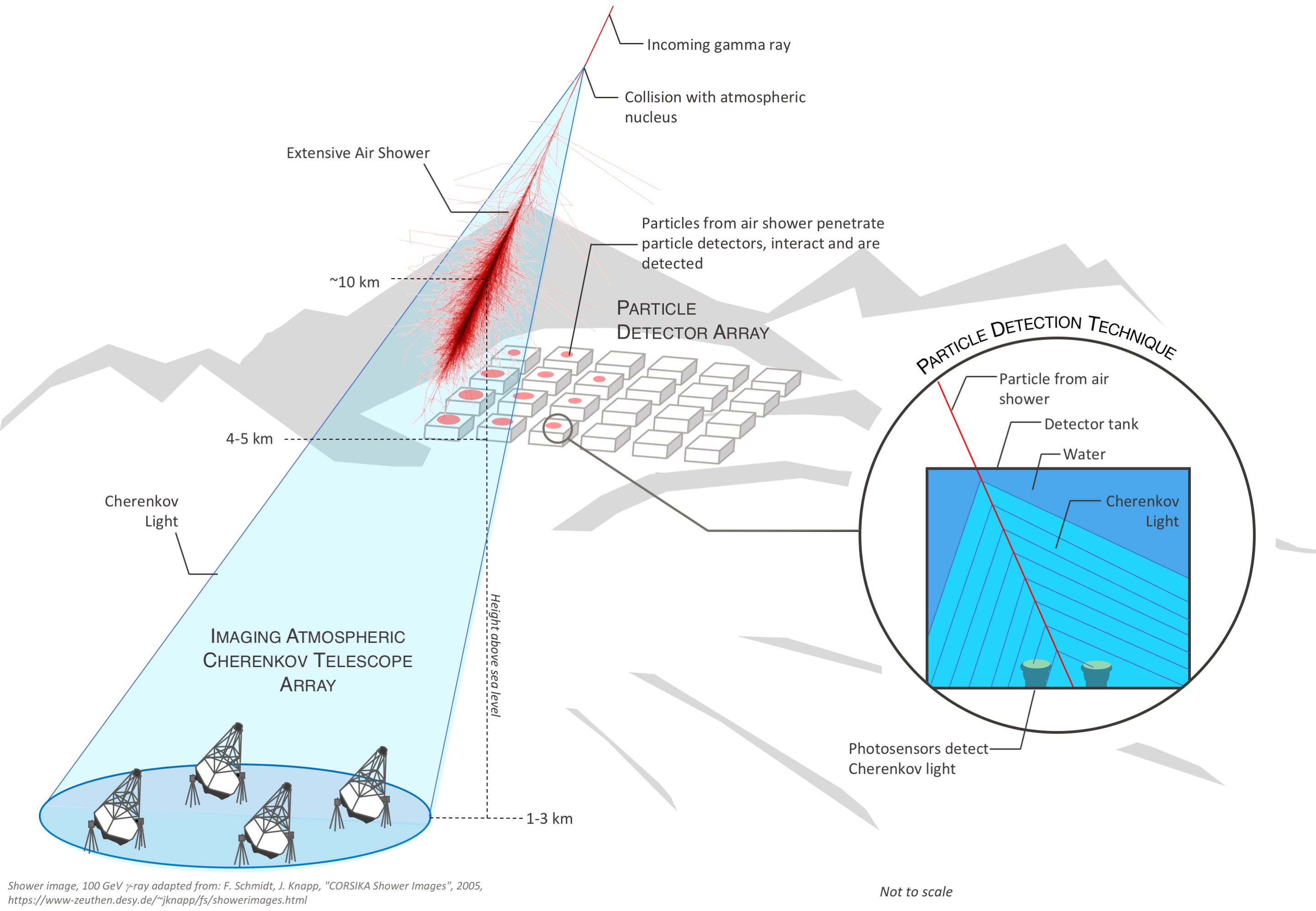}
    \caption{Schematic view of the future particle detector array SWGO, showing the detector layout and the detection technique, in comparison to a Cherenkov telescope array.}
    \label{fig:swgo}
\end{figure}

High-altitude water Cherenkov detectors are one of the most successful technologies to explore the gamma-ray sky from Earth. A future experiment of that kind will be the Southern Wide-Field Gamma-ray Observatory (SWGO), which is currently beeing developed by the SWGO collaboration \cite{SWGO}. The instrument will be sensitive to photons with energy from several tens of GeV to several hundreds of TeV, probing the very-high-energy (VHE) sky from the Southern Hemisphere and therefore complementing the HAWC and LHAASO experiments on the Northern Hemisphere. Its wide field of view will allow observing large regions of the sky, but will also provide the community with alerts of transient events in the VHE regime (see Fig.~\ref{fig:swgo} for details about the detection technique). The detection of Cherenkov light  produced in water by the particles of the gamma-ray showers, requires fast and efficient photodetectors and a reliable method of gamma/hadron separation, since the event rate of cosmic-ray showers is much larger than the event rate of gamma-ray showers. These characteristics, together with a low cost requirement for the components, are basic needs also for other experiments and use cases \cite{WOM}. In this proceedings we summarize our efforts to build and test a photo detector responding to all these requirements in the framework of SWGO. A special focus will be put on the efficiency and temporal performance of the WLS light traps, keeping in mind that such technology can be considered not only for the SWGO experiment but for many more applications in the future.

\section{Wavelength-shifting light traps - concept and motivation} \label{sec:description}

\begin{figure}[t]
    \centering
    \includegraphics[height=4cm]{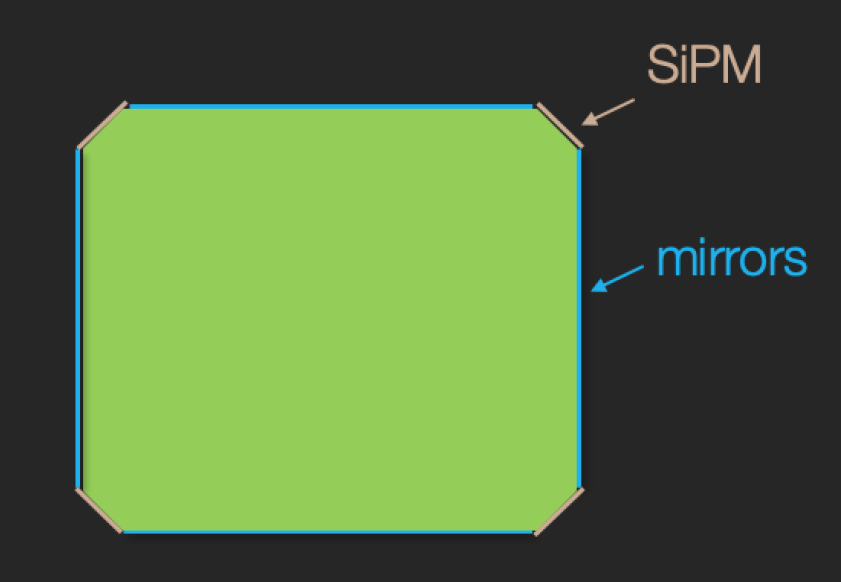}
    \includegraphics[height=4cm]{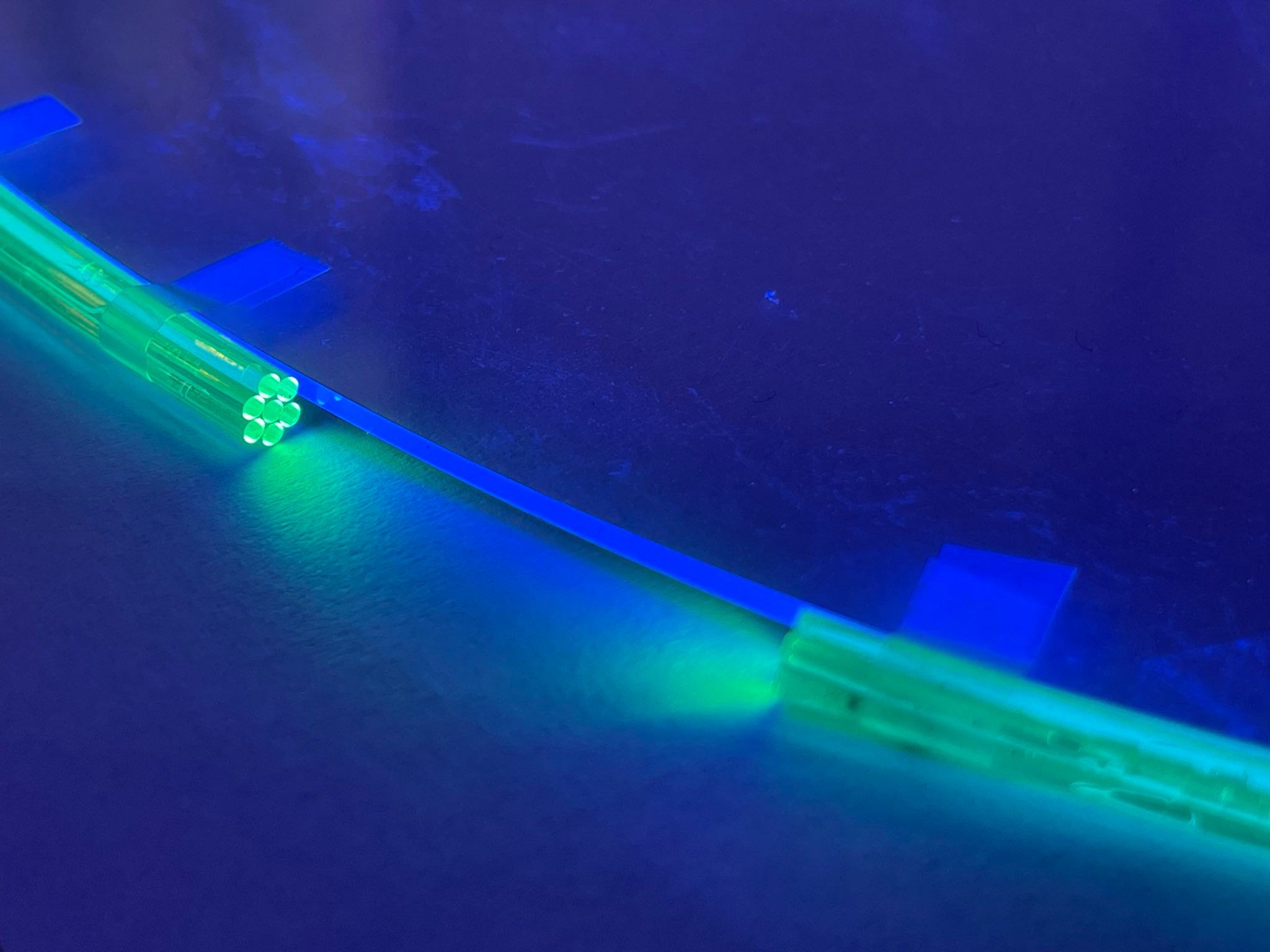}
    \caption{Single- and double-shift layout of a WLS light trap. Left: Scheme of the envisaged single-shift layout with a square WLS tile, enclosed by custom-made SiPM matches and reflecting material. Special SiPMs (1x30\,mm) with a shape similar to matches will be produced for that purpose. Right: Prototype of a double-shifting light trap with a WLS disk and fiber bundle.}
    \label{fig:designs}
\end{figure}

\begin{figure}[b]
    \centering
    \includegraphics[height=4cm]{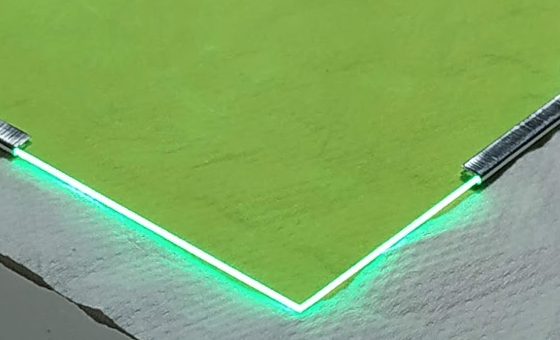}
    \hspace{1cm}
    \includegraphics[height=4cm]{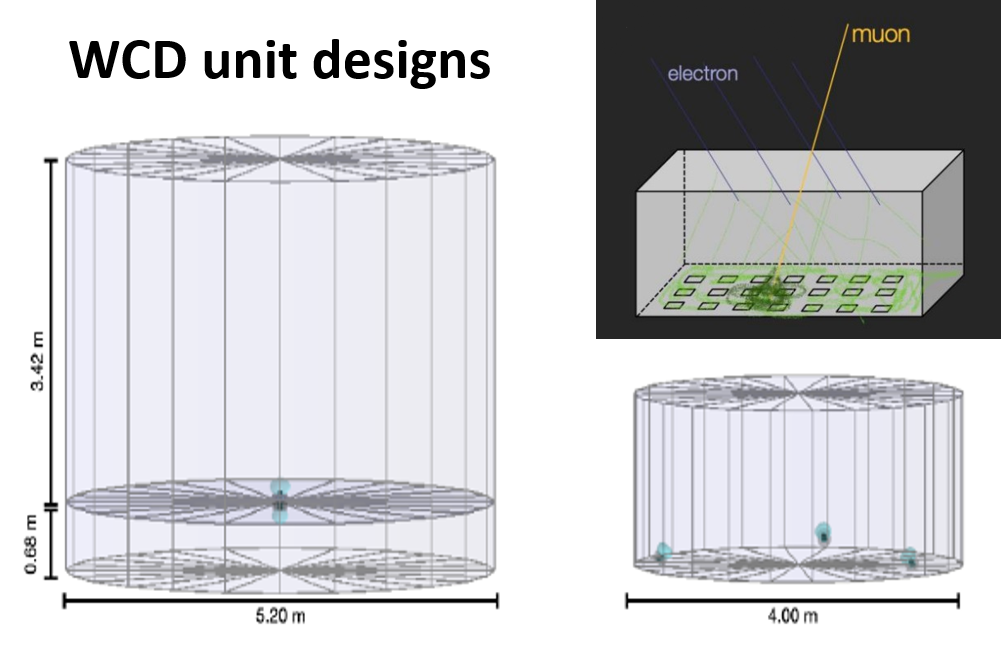}
    \caption{Left: Image depicting how photons hitting the WLS plate are absorbed, re-emitted and compressed to the small lateral surfaces by total internal reflection. Right: Concept of different WCD units, including a one-chamber tank with an array of light traps on the bottom (upper right) and two other designs considered for SWGO: a two-chamber tank with 1 PMT per chamber (left) and a one-chamber tank with multi-PMT readout (lower right).}
    \label{fig:motivation}
\end{figure}

We propose an alternative detector unit (compared to the current state-of-the-art photomultiplier tubes (PMTs)) called wavelength-shifting (WLS) light traps (see Fig.~\ref{fig:designs}). The name contains the two main effects which are exploited for an effective light detecting module. Primary Cherenkov light, emitted from particles in the UV band, is shifted to blue (and possibly re-shifted to green) through absorption and re-emission processes and a large percentage of the re-emitted light is trapped and guided towards the edges inside of the WLS material due to total internal reflection. The light trap compresses photons from a large entrance window (the surface of the plate) to the small lateral surfaces (see Fig.~\ref{fig:motivation}, left panel). At the edges of the material a Silicon photomultiplier (SiPM) detects the wavelength-shifted light. In this approach, several light traps could be arranged in an array at the bottom of each water Cherenkov detection unit (see Fig.~\ref{fig:motivation}, right panel). Muons should leave an inhomogeneous signal in the array of photo detection modules, electrons a more homogeneous signal. Therefore, this alternative layout of the WCD unit is expected to provide a good discrimination between gamma-ray and hadronic showers (to be shown by simulations). Additionally the design allows to reduce the height of the tank, the amount of water needed and therefore also the cost. 

We are considering two possible designs: A single-shift layout (see Fig.~\ref{fig:designs}, left panel), where custom made SiPMs (1x30\,mm long and thin, like matches) are positioned at the border of a square WLS plate (the residual edges are covered with some material of high reflectance in the wavelength range of interest). And a double-shift layout (see Fig.~\ref{fig:designs}, right panel), where a WLS disk is enclosed by a WLS fiber that re-shifts the light a second time and allows to have only two single SiPMs for light detection at both ends of the fiber. In both cases, the WLS disks can have large light-collecting surfaces ($\geq$ 500 $\text{cm}^{2}$) and still represent low cost components in view of mass production. The second option is less performant in terms of detection efficiency and time resolution, because the transitions between different materials make the system more lossy and the multiple shifts make it slower.

For our tests, WLS materials from the US company Eljen Technology (EJ-282 and EJ-286 tiles of different geometries, see \cite{Eljen}) and the Japanese company Kuraray (fibers of types YS-2, YS-4 and YS-6, see \cite{Kuraray} for the company's website and \cite{private}) were chosen. The differences in spectral absorption/emission peaks, absorption efficiency and temporal performance are due to the type, density and decay time of the dopant molecule (see Tab.~\ref{tab:materials} for details). The materials were chosen such to optimize the matching of material absorption/emission peak with the Cherenkov spectrum and with the photon detection efficiency (PDE) of the SiPMs.

In addition, Hamamatsu SiPMs (S14160/S14161 series, see \cite{Hamamatsu}) with a size of 3x3\,$\text{mm}^2$ are used for the light trap detecting modules. As Geiger-mode avalanche photodiodes (G-APD) the SiPMs have single photon sensitivity and a high PDE ranging from 30\,\% to 50\,\% in the wavelength range from 300\,nm to 600\,nm ($\approx$ 50\,\% at the emission peak of the WLS materials). 

\begin{table}[h]
    \centering
    \begin{tabular}{l|lllll}
         & EJ-282 & EJ-286 & YS-2 & YS-4 & YS-6 \\
         \hline
        Absorption peak (nm) & 390 & 355 & 422 & 420 & 414 \\
        Emission peak (nm) & 481 & 425 & 474 & 470 & 462 \\
        Decay time (ns) & 1.9 & 1.2 & 3.2 & 1.4 & 1.3 \\
    \end{tabular}
    \caption{Characteristics of the applied WLS materials from Eljen Technology \cite{Eljen} and Kuraray \cite{Kuraray}. All of these materials have a refractive index of around $\text{n}=1.6$ and consist of a similar polymer base material.}
    \label{tab:materials}
\end{table}

\section{Spectral measurements} \label{sec:spectra}

\begin{figure}[ht]
    \centering
    \includegraphics[width=0.49\textwidth]{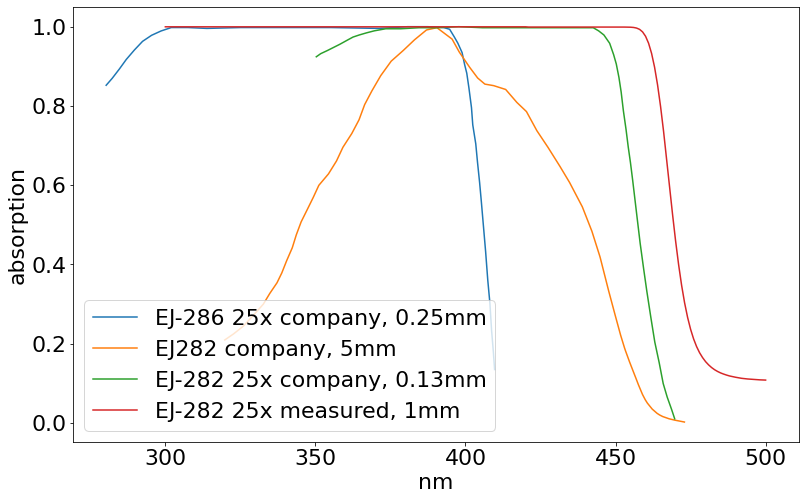}
    \includegraphics[width=0.49\textwidth]{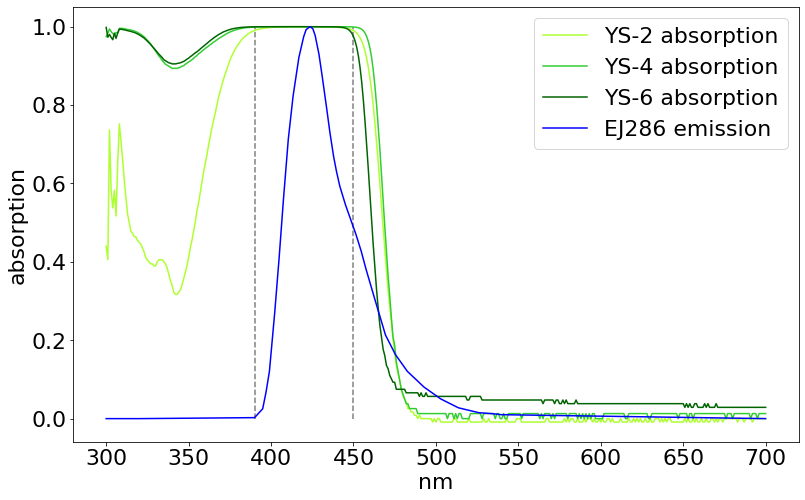}
    \caption{Left: Measured absorption spectrum of EJ-282 (25 times standard dopant concentration), compared to spectra of EJ-282 and EJ-286, provided by the company. The numbers in mm in the legend indicate the thickness of the tile. Note that the absorption peaks are not visible for high dopant concentrations because the absorption is maximized in a large wavelength range. Right: Confront of spectral emission curve of the EJ-286 disk \cite{Eljen} and absorption curves for the YS-series fibers from Kuraray \cite{private}, in a configuration of a 7-fiber bundle. The mismatch between these curves causes a loss of several percent efficiency. The grey dashed lines mark the wavelength range in which the 7-fiber bundles of YS-2, YS-4 or YS-6 fibers totally absorb the light.}
    \label{fig:spectra}
\end{figure}

To verify and confirm the spectral absorption and emission curves of the materials provided by the companies, the curves were measured in the laboratory using small samples of the WLS materials. The emission and absorption curves we measured were all very similar to the ones provided by the company. Apart from the consistency with the companies measurements, it was important to confirm the assumption of the total absorption by the WLS tiles, given its impact on the maximum achievable overall efficiency. It can be concluded that for the EJ-286 and EJ-282 disks of 1\,mm thickness and 25 times the standard concentration of the dopant, total absorption in the concerned wavelength range is fulfilled (see Fig.~\ref{fig:spectra}, left panel). The spectral absorption curves for the fibers of type YS-2, YS-4 and YS-6 (provided by the company \cite{private}) confirm the total absorption in the wavelength range from 400\,nm to 450\,nm for a bundle of 7 fibers (see Fig.~\ref{fig:spectra}, right panel).

\section{Light detection efficiency} \label{sec:eff}

 One critical property to be characterized is the light detection efficiency of the WLS light traps. We did this with a simple experimental setup inside a black container, using a pulsed LED (for the emission spectrum, see Fig.~\ref{fig:mismatch_fiber_lamp_PDE} below) as light source and an oscilloscope for the SiPM readout. 

\subsection{Single-shift layout}
The left panel of Fig.~\ref{fig:designs} shows the setup envisaged for the single-shift layout. Assuming to have 4 SiPM matches (3\,cm length), attached along the borders of a WLS tile (EJ-286, 15x15\,$\text{cm}^2$), we expect the following losses: 

\begin{itemize}
\begin{spacing}{0.1} 
    \item disk absorption efficiency: $\approx$ 100\,\% (see Sec.~\ref{sec:spectra})
    \item disk re-emission efficiency (loss to phonons): 92\,\% for EJ-286, 90\,\% lower limit used (see data sheets \cite{Eljen})
    \item disk trapping efficiency in air and water: $\approx$ 77\,\% and $\approx$ 54\,\% (as calculated)
    \item losses through re-absorption inside the tile (small because attenuation length much larger than dimensions of the disk)
    \item losses from imperfect reflection on borders (considered reflecting material has > 97\,\% reflectivity between 400\,nm and 800\,nm, mean number of reflections undergone by a photon before detection unkown) and losses at the transition between disk and SiPM (unknown) 
\end{spacing}
\end{itemize}

Using the information above, upper limits for a partial efficiency of the system (only including the effects of absorption, re-emission and trapping but not reflection and detection by the SiPM) can be estimated as $0.92 \cdot 0.77 \approx 0.71 = 71$\,\% and 50\,\% for EJ-286 in air and for EJ-286 in water, respectively. Note, that the calculation is based on the assumption that all the compressed light that exits the lateral surfaces of the tile is detected. The total efficiency will be reduced due to the fact that only part of the border is covered with SiPM matches and it will depend on the quality of the reflective material along the residual border. We repeatedly measured the above partial efficiency of the single-shift layout in air. Taking the weighted mean of all the measurements and calculating the weighted uncertainty gives an efficiency of ($69.2 \pm 2.5$)\,\%, which is consistent with the theoretically calculated partial efficiency of 71\,\% for EJ-286 in air. For the complete setup, in which losses due to imperfect reflections and the PDE of the SiPM play a role, we expect to reach at least 10\,\% and possibly higher total efficiencies in the single-shift layout, both in air and water.

\subsection{Double-shift layout}
The right panel of Fig.~\ref{fig:designs} shows a possible setup envisaged for the double-shift layout. It consists of a WLS disk and either a single fiber or a bundle of 3 or 7 fibers, enclosing the border of the disk. At the ends of the fiber (bundle) the light is detected by two single SiPMs of 3x3\,$\text{mm}^2$ size. Assuming to have a 7-fiber bundle enclosing a WLS disk of EJ-286 material (Diameter: 25\,cm, Thickness: 1\,mm), we expect the following losses: 

\begin{itemize}
\begin{spacing}{0.1} 
    \item all the effects listed for the single-shift layout except for the last point (here using re-emission efficiency of EJ-286: 92\,\%)
    \item losses at the transition between disk and fiber: expected to be non-negligible but unknown, at least $\approx$ 4\,\% because of losses in a small angle between the fiber ends (see Fig.~\ref{fig:designs}, right panel), estimated with 3\,cm gap along the perimeter
    \item absorption efficiency of fiber(s): depends on number of fibers and the overlap of disk emission spectrum and fiber absorption spectrum (see Fig.~\ref{fig:spectra}, right panel)
    \item fiber re-emission efficiency (loss to phonons): assuming $\approx$ 90\,\%
    \item fiber trapping efficiency in air: $\approx$ 37\,\% (as calculated)
    \item losses through re-absorption inside the fiber (small because the attenuation length is much larger than the dimensions of the disk, see \cite{Kuraray})
    \item losses at the transition between fiber and SiPM (unknown)
\end{spacing}
\end{itemize}

\begin{figure}[h]
    \centering
    \includegraphics[width=0.55\textwidth]{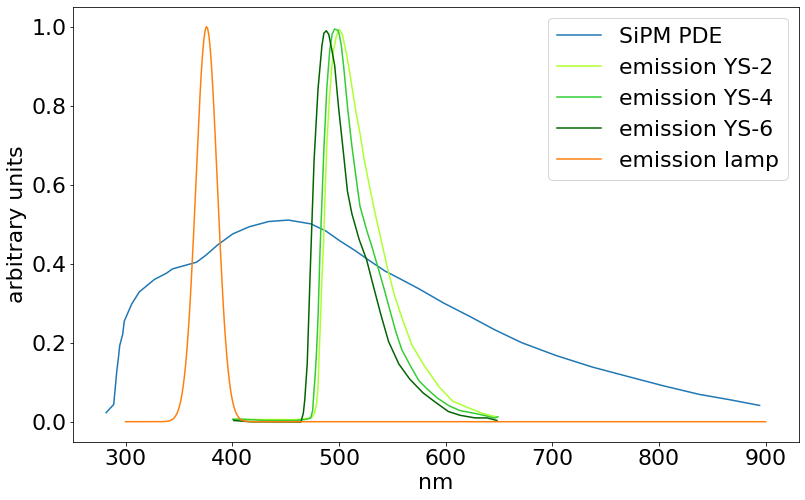}
    \caption{Comparison of the emission spectra of the different WLS fibers and the LEDs emission spectrum as well as the PDE of the SiPM. The mismatch between the emission spectra and PDE causes a change in the measured efficiency and because this mismatch is not equally strong for the fiber emission and the LED spectrum, a systematic uncertainty is introduced in the measurement.}
    \label{fig:mismatch_fiber_lamp_PDE}
\end{figure}

Using the information above, we estimated upper limits (ULs) for the overall efficiency of a double-shift layout in different configurations. The total theoretical efficiencies range between 17\,\% and 20\,\% in air, depending on the fiber type and the number of fibers used. In water these efficiencies are reduced to 5\,\% - 6\,\%. Note, that we introduce an unavoidable but rather small systematic error due to a mismatch between the emission spectra of the fibers/the pulsed LED and the PDE of the SiPM (see Fig.~\ref{fig:mismatch_fiber_lamp_PDE}). If the PDE was constant for all wavelengths in our measurement, the effect would cancel out, because it affects both the measurement of intensity with the light trap and the direct reference measurement, of which the ratio is computed. In air, we measured a maximum efficiency of 3\,\%, for a double-shift layout with 7 YS-2 fibers, optical coupling (OC) and reflecting foil around the fibers, and we confirmed the result by repeated measurements. Comparing measurements with and without OC or reflecting foil, we conclude that both features definitely increase the efficiency of the system. Considering the above, a reasonable estimate of the systematic uncertainty on the measurements is at least $\pm$ 1.0\,\%. Therefore, we conclude that this result is only roughly consistent with our expectation. We plan to further improve our prototypes and reduce systematic losses, hoping to reach higher efficiencies, closer to our theoretical predictions for the efficiency in air.

\section{Temporal performance of single- and double-shift layout}

\begin{figure}[h]
    \centering
    \includegraphics[width=0.55\textwidth]{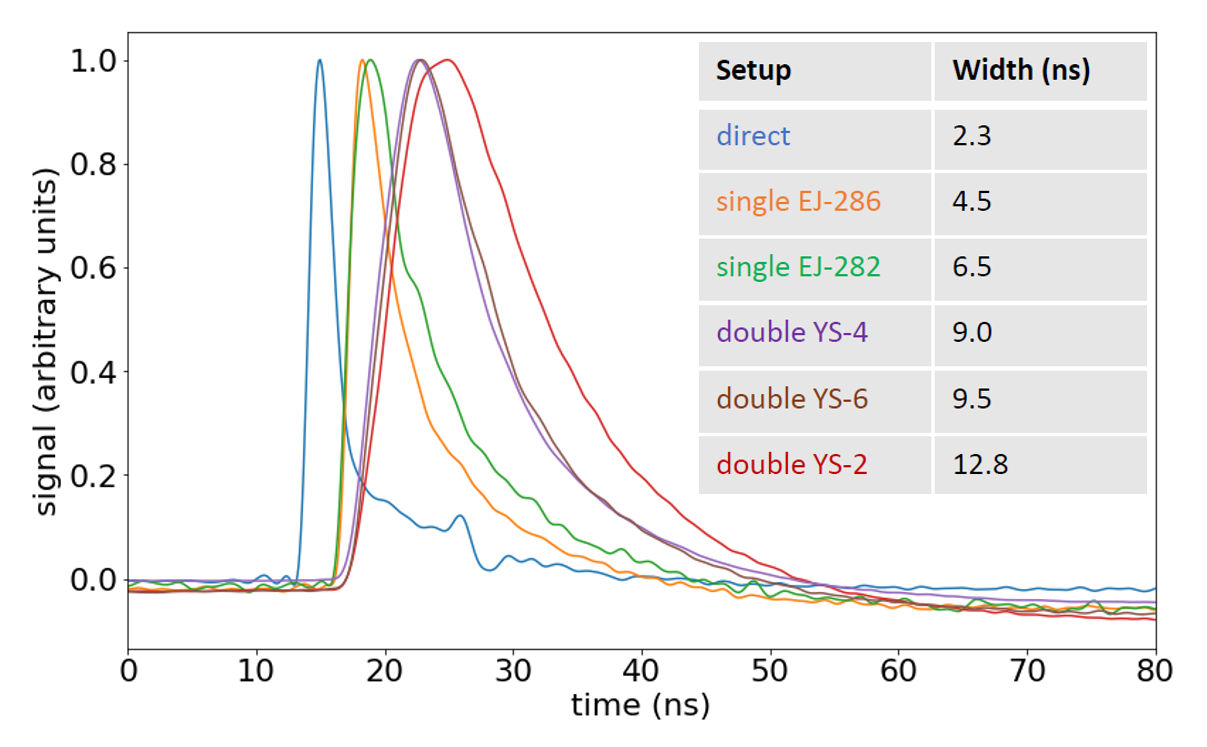}
    \caption{Measured pulse shapes of the signal for different light-trap designs. From narrowest to widest pulse: Signal measured directly with the SiPM, the single-shift layout with EJ-286 and EJ-282, the double-shift layout (always with EJ-286 disk) with fiber types YS-4 and YS-6 and the double-shift layout with fiber type YS-2. Decay times of all materials can be found in Tab.~\ref{tab:materials}.}
    \label{fig:pulse_shape}
\end{figure}

The temporal performance of all different layouts was measured by comparing the characteristics of the signal shape from the single- and double-shift design with a direct measurement of the pulse shape (limited by the electronics and SiPM readout, not by the pulse width of the LED itself) as reference (see Fig.~\ref{fig:pulse_shape}). As can be seen from the plot and the table the temporal performance of the light traps is good. Even for the double-shift layout with the fibers YS-4 and YS-6, the width of the signal pulse is less than 10\,ns. For the single-shift layout the width varies between 4\,ns and 6\,ns, depending on the molecule decay time. Compared to PMTs the time resolution is slower by a factor of two for the single-shift layout and by a factor of three to six for the double-shift layout, but should still be enough for a good direction reconstruction of the showers, observed by SWGO. We expect that having many light traps in one detection unit and taking a mean of the signal can reduce this uncertainty.

\section{Cost estimate}
For a single-shift design with 4 SiPM matches attached along the borders of a WLS tile, the following cost can be estimated: We paid 250\,\euro{} per WLS plate and 25\,\euro{} per SiPM but considering a mass production for 5000 water Cherenkov detector units the cost will be strongly reduced. Therefore, we assume 100\,\euro{} and 10\,\euro{} for the cost of one tile and one SiPM match, respectively. Assuming to put 4 SiPMs for the readout of each light trap and 6 light traps per detection unit, this would mean a total cost of 4.2 Million\,\euro{}. This corresponds to 40\,\% of the estimated cost, given for the scenario of PMTs for SWGO \cite{SWGO}. It is only a rough preliminary estimate (without consideration of possible additional costs, e.\,g. for reflecting elements along the border, electronics and structural parts for the integration into the water tank) but we take this as a tentative indication for the feasibility of the light-trap solution in terms of cost. In terms of efficiency however, the competitiveness still has to be evaluated in detail and we provided information about it in this contribution.  

\section{Conclusion and Outlook}
We have presented an alternative photo detection module called wavelength-shifting light trap, using low cost WLS materials and SiPMs for light detection. Results concerning the light detection efficiency and the temporal performance of different possible layouts show that WLS light traps are efficient, fast and cheap and a promising technology to be developed, improved and possibly applied for SWGO or other future Cherenkov experiments/applications.~Our currently favored design is the single-shift layout, which is more efficient, more robust and less exposed to systematic uncertainties, introducing less potential sources of loss in the system. Future work on this project will be focused on the improvement of efficiency (also trying other materials), construction of a first complete light trap in the single-shift layout, using SiPM matches, and the gradual development of additional components, needed to integrate a light trap inside of a tank.

\small
\acknowledgments
We would like to thank the Department of Physics and Astronomy of the University of Padua and the INFN section of Padua for the funding of our R\&D activities, the chemist Sara Carturan (Univ.~and INFN Padua) and the technician Luca Silvestrin (Univ.~and INFN Padua) for their advice and support, and the companies Kuraray, Eljen Technology, Hamamatsu and FBK for providing the newest technological components for our light traps and for the private communication about product characteristics.

%
%
%


\begin{thebibliography}{99}

\bibitem{SWGO}
P. Huentemeyer and the SWGO Collaboration,
\emph{The Southern Wide-Field Gamma-Ray Observatory (SWGO): A Next-Generation Ground-Based Survey Instrument for VHE Gamma-Ray Astronomy}, 
Astro2020 APC White Paper,
\href{https://arxiv.org/abs/1907.07737} {arXiv:1907.07737}.

\bibitem{WOM}
L. Schulte et al.,
\emph{A large-area single photon sensor employing wavelength-shifting and lightguiding technology}, 
ICRC 2013, Rio de Janeiro,
\href{https://arxiv.org/abs/1307.6713v1} {arXiv:1307.6713v1}.

\bibitem{Eljen}
Eljen Technology,
\emph{Wavelength Shifters},
\href{https://eljentechnology.com/products/wavelength-shifting-plastics} {https://eljentechnology.com/products/wavelength-shifting-plastics}.

\bibitem{Kuraray}
Kuraray,
\emph{Plastic Scintillating Fibers (PSF)},
\href{https://www.kuraray.com/products/psf} {https://www.kuraray.com/products/psf}.

\bibitem{private}
Private Communication with companies.

\bibitem{Hamamatsu}
Hamamatsu,
\emph{MPPC: S14160/S14161 series},
\href{https://www.hamamatsu.com/content/dam/hamamatsu-photonics/sites/documents/99_SALES_LIBRARY/ssd/s14160_s14161_series_kapd1064e.pdf} {https://www.hamamatsu.com/content/dam/hamamatsu-photonics/sites/documents/99\_SALES\_LIBRARY/ssd/s14160\_s14161\_series\_kapd1064e.pdf}.


\end{thebibliography}
\end{document}